\documentclass[twocolumn]{revtex4}

\begin{document}
\title{
Comment to ``Nonextensive Thermodynamics and Glassy Behaviour in
Hamiltonian Systems" by A. Rapisarda and A. Pluchino, Europhysics News
36, 202 (2005).}

\author{
F. Bouchet ({\em INLN, Nice}, France, {\tt Freddy.Bouchet@inln.cnrs.fr}),\\
T. Dauxois ({\em ENS Lyon}, France, {\tt Thierry.Dauxois@ens-lyon.fr}) and\\ 
S. Ruffo   ({\em Florence University}, Italy, {\tt Stefano.Ruffo@unifi.it}) }
\date{\today} 

\maketitle

The dynamics of the Hamiltonian Mean-Field (HMF) model~\cite{book} shows
many intriguing non-equilibrium behaviors. In particular, it has been
reported several times that the system gets stuck into {\em
quasi-stationary states} (QSS), whose lifetime increases with system
size. As correctly pointed out by C. Tsallis and coworkers (see
e.g. Refs.~\cite{Tsallis}), the presence of such non-equilibrium
states is tightly linked to the fact that the infinite time limit and
the thermodynamic limit do not commute in systems with long-range
interactions. However, contrary to what is claimed in
Ref.~\cite{Rapisarda}, the non-extensive statistics approach does not
convincingly ``explain'' any of these non-equilibrium behaviors.

Two main quantities have been tested up to now: velocity distribution
functions and correlation functions. In Ref.~\cite{latora}, the
authors fit single particle velocity distribution functions in QSS
using $q$-exponentials. They obtain a value of the index $q=7$. In
Ref.~\cite{Rapisarda}, an analogous fit of correlation functions with
$q$-exponentials gives values of $q$ between 1.1 and 1.5. It is
questionable that different values of $q$ are used for the same model
and the same physical conditions.

The fact of being in a non-equilibrium state could in principle allow
the use of an entropy other than Boltzmann-Gibbs. However, there is up
to now not a single paper which gives a rigorous justification
of the use of non-extensive entropy for
the HMF model. Hence, there is no compelling reason of using $q$-exponentials as
a fitting function. 


A general alternative approach has been introduced to {\em explain} the
presence of QSS in systems with long-range interactions.  This
approach begins by performing first the thermodynamic limit and then
looking at the time evolution.  This procedure amounts to associate to the HMF model
appropriate Vlasov and kinetic equations.  This method is fully
predictive and has been extensively exploited in Ref.~\cite{Yoshi} to
obtain the Vlasov equation predictions for the HMF model. 

Restricting to {\em homogeneous} QSS, velocity
distribution functions of QSS have been analysed, reaching the
conclusion that they cannot be fitted by $q$-exponentials.  This
conclusion has not been questioned so far in the literature.
Moreover, the kinetic approach also allows to
derive properties of the correlation functions, deducing them directly
from the HMF model~\cite{freddy}.  Such homogeneous states are of
paramount importance, since they appear to be ``attractive'' for a
large class of initial conditions. For instance, it can be shown that
the plateau values of the magnetization $M_0$ shown in Fig.~1 of
Ref.~\cite{Rapisarda}, all converge to $M_0=0$ when $N$ increases,
which is a distinctive sign of homogeneity.

The Vlasov equation approach is just in a beginning stage. However,
the already existing results are encouraging and we believe that the
difficulty of treating inhomogeneous QSS is of technical nature.
This problem will be solved in the near future.

Hence, the conclusion of Ref.~\cite{Rapisarda}: ``However the actual state of
the art favours the application of non-extensive thermostatistics to
explain most of the anomalies observed in the QSS regime'' is 
highly questionable. 

As a final remark, we think that, as physicists, we should pay great
attention to the difference between ``fitting'' and ``explaining''.

\end{document}